\documentclass[reprint,superscriptaddress,amsmath,amssymb,twocolumn,amsfonts,floatfix, longbibliography]{revtex4-2}
\usepackage{graphicx}
\usepackage{epstopdf}
\usepackage{amsbsy}
\usepackage{gensymb}

\usepackage[dvipsnames]{xcolor}
\definecolor{deepfuchsia}{rgb}{0.76, 0.33, 0.76}
\definecolor{electricpurple}{rgb}{0.75, 0.0, 1.0}
\usepackage[colorlinks=true,linktoc=page,linkcolor=blue,citecolor=magenta,urlcolor=electricpurple]{hyperref}

\usepackage{orcidlink}
\usepackage{float}

\newcommand{\beq}{\begin{equation}}
\newcommand{\eeq}{\end{equation}}
\newcommand{\bea}{\begin{eqnarray}}
\newcommand{\eea}{\end{eqnarray}}

\newcommand{\fig}[1]{Fig.~\ref{#1}}

\usepackage{mathtools}

\DeclarePairedDelimiterX\braket[2]{\langle}{\rangle}{#1 \delimsize\vert #2}

\begin{document}
\title{Nonunitary triplet superconductivity in the $\mathbb{Z}_2$ topological metal SrPd$_{2}$As$_{2}$}

\author{Aarti}\thanks{These authors contributed equally to this work}
\affiliation{Department of Physics, University of Petroleum and Energy Studies, Dehradun, Uttarakhand, 248007, India}
\author{{Dibyendu Samanta}\,\orcidlink{0009-0004-3022-7633}}
\thanks{These authors contributed equally to this work}
\affiliation{Department of Physics, Indian Institute of Technology, Kanpur 208016, India}
\author{Kartik Panda}
\affiliation{Department of Physics, Ariel University, Ariel 40700, Israel}
\author{Devashibhai Adroja}
\altaffiliation{devashibhai.adroja@stfc.ac.uk}
\affiliation{ISIS Facility, Rutherford Appleton Laboratory, Chilton, Didcot, Oxon, OX11 0QX, United Kingdom}
\affiliation {\mbox{Highly Correlated Matter Research Group, Physics Department, University of Johannesburg,} P.O. Box 524, Auckland Park 2006, South Africa}
\author{Daloo Ram}
\affiliation {Department of Physics, Indian Institute of Technology, Kanpur 208016, India}
\affiliation {School of Advanced Materials and Chemistry and Physics of Materials Unit, Jawaharlal Nehru Centre for Advanced Scientific Research, Jakkur, Bangalore 560064, India}
\author{Zakir Hossain}
\affiliation {Department of Physics, Indian Institute of Technology, Kanpur 208016, India}
\author{Rhea Stewart}
\affiliation{ISIS Facility, Rutherford Appleton Laboratory, Chilton, Didcot, Oxon, OX11 0QX, United Kingdom}
\author{Adrian Hillier}
\affiliation{ISIS Facility, Rutherford Appleton Laboratory, Chilton, Didcot, Oxon, OX11 0QX, United Kingdom}

\author{Amitava Bhattacharyya}
\affiliation{Department of Physics, Ramakrishna Mission Vivekananda Educational and Research Institute, Belur Math, Howrah 711202, West Bengal, India}
\author{Samar Layek}
\affiliation{Department of Physics, University of Petroleum and Energy Studies, Dehradun, Uttarakhand, 248007, India}

\author{{Sudeep Kumar Ghosh}\,\orcidlink{0000-0002-3646-0629}}
\email[]{skghosh@iitk.ac.in}
\affiliation{Department of Physics, Indian Institute of Technology, Kanpur 208016, India}

\author{{Vivek Kumar Anand}\,
\orcidlink{0000-0003-2023-7040}}
\email{vivekkranand@gmail.com}
\affiliation{Department of Physics, National Institute of Technology Agartala, Tripura 799046, India}
\affiliation{Department of Mathematics and Physics, University of Stavanger, 4036 Stavanger, Norway}

\date{\today}

\begin{abstract}
In $\mathbb{Z}_2$ topological metals, nontrivial band topology and strong spin–orbit coupling (SOC) impose symmetry constraints that can stabilize unconventional superconducting states, even when thermodynamic probes indicate an isotropic gap. Here, we investigate the superconducting ground state of such a material, SrPd$_2$As$_2$, using muon spin rotation and relaxation ($\mu$SR), first-principles calculations, and Ginzburg-Landau analysis. Transverse-field $\mu$SR indicates a fully gapped superconducting state below $T_c \approx 0.94$ K, while zero-field $\mu$SR detects spontaneous internal magnetic fields below $T_c$, establishing time-reversal symmetry (TRS) breaking. Electronic structure calculations identify SrPd$_2$As$_2$ as a $\mathbb{Z}_2$ topological metal with surface states crossing the Fermi level. Standard anisotropic Migdal-Eliashberg calculations predict a nodal gap and overestimate $T_c$, indicating that a purely phonon-mediated pairing mechanism is insufficient. We resolve this apparent contradiction by showing that the interplay of SOC, tetragonal symmetry, and an open Fermi surface topology stabilizes a nonunitary triplet superconducting state whose symmetry-imposed nodes lie in momentum-space regions devoid of electronic states. This yields a fully gapped thermodynamic response while naturally breaking TRS. Our results establish SrPd$_2$As$_2$ as a clean platform for bulk nonunitary triplet pairing and a promising candidate for intrinsic topological superconductivity.

\end{abstract}

\maketitle


\section{Introduction}
The discovery of topological metals~\cite{Armitage2018,Lv2021} has introduced a class of systems in which relativistic band structure and crystal symmetry can qualitatively influence superconducting behavior. In several such materials, superconductivity is accompanied by the spontaneous breaking of time-reversal symmetry (TRS), signaled by the emergence of weak internal magnetic fields below the transition temperature ($T_c$)~\cite{ghosh2021Recent}. Muon spin rotation and relaxation ($\mu$SR) provides a sensitive probe of these fields and has established TRS-breaking superconductivity in a range of systems, including Sr$_2$RuO$_4$~\cite{luke1998time,Maeno2024}, LaPt$_3$P~\cite{Biswas2021chiral}, SrPtAs~\cite{SrPtAs}, and nonunitary triplet superconductors such as LaNiC$_2$~\cite{hillier2009evidence} and LaNiGa$_2$~\cite{Hillier2012nonunitary,Ghosh2020b,badger2022dirac}. Despite these advances, the microscopic origin of TRS breaking remains strongly material-dependent. This is particularly true in systems that exhibit a fully gapped thermodynamic response, where conventional expectations favor a real, single-component order parameter, thus creating a contradiction between experimental observations and standard pairing models~\cite{bhattacharyya2018unconventional,bhattacharyya2015unconventional,bhattacharyya2015broken,singh2020time,mayoh2021evidence,barker2015unconventional,bhattacharyya2022nodeless,shang2020time,anand2023time, aarti2024time}.

A natural setting to address this issue is provided by materials in which superconductivity coexists with nontrivial band topology and strong spin-orbit coupling (SOC). In such systems, symmetry and electronic structure can restrict the allowed pairing states and enable unconventional superconducting ground states~\cite{Meena2025superconductivity,Meena2025nonsym,Yadav2024}. Recent studies have reported TRS breaking superconductivity in several topological semimetals~\cite{Shang2022Weyl,Sajilesh2025time,Ghosh2022Dirac}; however, in many cases, additional factors such as geometric frustration or competing orders complicate the identification of the underlying mechanism~\cite{Neupert2022charge}. In this context, the 122-family of Pd-pnictides~\cite{anand2014superconductivity, anand2013superconducting, anand2023time, aarti2024time}, which crystallize in the high-symmetry ThCr$_2$Si$_2$-type structure, provides a chemically clean and symmorphic platform to isolate these effects. Previous measurements on single crystal SrPd$_2$As$_2$ indicate bulk superconductivity below $T_{c} = 0.92(5)$ K with signatures suggestive of unconventional behavior~\cite{anand2013superconducting}, yet the nature of its pairing state remains unresolved. This raises a central question: can SrPd$_2$As$_2$ host a TRS-breaking superconducting state that appears fully gapped in thermodynamic probes, and if so, what microscopic mechanism reconciles these observations?

In this work, we address this question using $\mu$SR measurements combined with first-principles calculations and Ginzburg-Landau symmetry analysis. Transverse-field $\mu$SR establishes a fully gapped superconducting state below $T_c \approx 0.94$~K, while zero-field measurements reveal spontaneous internal magnetic fields, demonstrating TRS breaking in SrPd$_2$As$_2$. We show that conventional phonon-mediated Migdal-Eliashberg theory fails to capture this behavior, instead predicting a nodal gap and overestimating $T_c$. In contrast, electronic structure calculations reveal that SrPd$_2$As$2$ is a $\mathbb{Z}_2$ topological metal with surface states at the Fermi level. By analyzing the symmetry-allowed pairing channels, we demonstrate that the interplay of SOC, symmorphic $D_{4h}$ symmetry, and the specific Fermi surface topology stabilizes a nonunitary triplet superconducting state. Crucially, the symmetry-imposed nodes of this state lie in regions of momentum space devoid of electronic states, yielding a fully gapped thermodynamic response while naturally breaking TRS. These results establish SrPd$_2$As$_2$ as a clean platform for studying nonunitary triplet pairing in topological metals and provide a concrete mechanism for reconciling full-gap behavior with TRS breaking.

\section{Methods}

\subsection{Synthesis and Characterization of samples}

Polycrystalline samples of SrPd$_2$As$_2$ were synthesized using a conventional solid-state reaction technique, following procedures similar to those reported for CaPd$_2$As$_2$~\cite{aarti2024time} and CaPd$_2$Ge$_2$~\cite{anand2023time}. Stoichiometric amounts of high-purity elemental precursors (Sr: 99.9\%, Pd: 99.95\%, and As: 99.999\%) were used, with Sr taken in the form of fine pieces and Pd and As as powders. The elements were thoroughly mixed, pressed into pellets which were heat-treated at 800$^\circ$C for 30 hours after sealing in an evacuated quartz tube. The resulting material was then ground, pelletized, resealed in a quartz tube, and annealed at 900$^\circ$C for 72 hours. This grinding and annealing cycle was repeated, followed by an additional heat treatment at 900$^\circ$C for 72 hours to improve the phase formation. The powder X-ray diffraction (XRD) (using Cu-$K_\alpha$ radiation) confirmed the phase purity of the synthesized SrPd$_2$As$_2$ sample and also provided the crystallographic information. The room temperature XRD pattern is shown in Appendix (Fig.~\ref{fig:XRD-Fig})

\subsection{$\mu$SR Experiments}

Muon spin rotation/relaxation ($\mu$SR) measurements were performed to investigate the superconducting gap structure and to test for possible time-reversal symmetry (TRS) breaking in SrPd$_2$As$_2$. The experiments were carried out using the MuSR spectrometer at the ISIS Facility, Didcot, UK. The sample mount was prepared by applying diluted GE varnish to the powdered SrPd$_2$As$_2$ sample held on a high-purity silver plate (Ag, 99.999\%) and wrapped with a thin silver foil. The ZF-$\mu$SR measurements were performed in zero field (ZF) using an active field-compensation system that suppresses stray magnetic fields to the microtesla level. The ZF-$\mu$SR measurements were performed at several temperatures over $0.07~\rm{K} \leq T \leq 2$~K by cooling the sample in a dilution refrigerator. The longitudinal-field (LF) $\mu$SR measurements were performed with longitudinally applied magnetic fields at 0.07~K\@. Transverse-field (TF) $\mu$SR measurements were carried out in the temperature range $0.07~\rm{K} \leq T \leq 0.8$~K under applied transverse magnetic fields of $H = 5$, 7.5, 10, 20, and 30~mT. The ZF-, LF- and TF-$\mu$SR data were analyzed using the WiMDA software package~\cite{pratt2000wimda}.

\subsection{Electronic band structure, topology, and phonons}

Electronic band structure calculations were performed within density functional theory (DFT) using the \textsc{QUANTUM ESPRESSO} package~\cite{Giannozzi_2009,Giannozzi_2017,Giannozzi_2020}. The generalized gradient approximation (GGA) with the Perdew--Burke--Ernzerhof (PBE) exchange-correlation functional~\cite{Perdew_1996} and projector augmented-wave (PAW) pseudopotentials were employed. A plane-wave kinetic energy cutoff of $80$~Ry and a $\Gamma$-centered $12 \times 12 \times 10$ $k$-point mesh were used for bulk Brillouin-zone sampling. Experimental lattice parameters and atomic positions obtained from Rietveld refinement were used as input. Maximally localized Wannier functions were constructed using the \textsc{WANNIER90} package~\cite{Pizzi2020}, with initial projections chosen as Sr-$4p$, Pd-$4d$/$4p$, and As-$4s$/$4p$ orbitals. The resulting tight-binding Hamiltonian was used to compute Fermi surfaces and surface spectral functions for a semi-infinite SrPd$_2$As$_2$ slab via the iterative Green's function method implemented in \textsc{WannierTools}~\cite{Wu2018}.

\begin{figure*} [!htb] 
\includegraphics[width=0.95\textwidth]{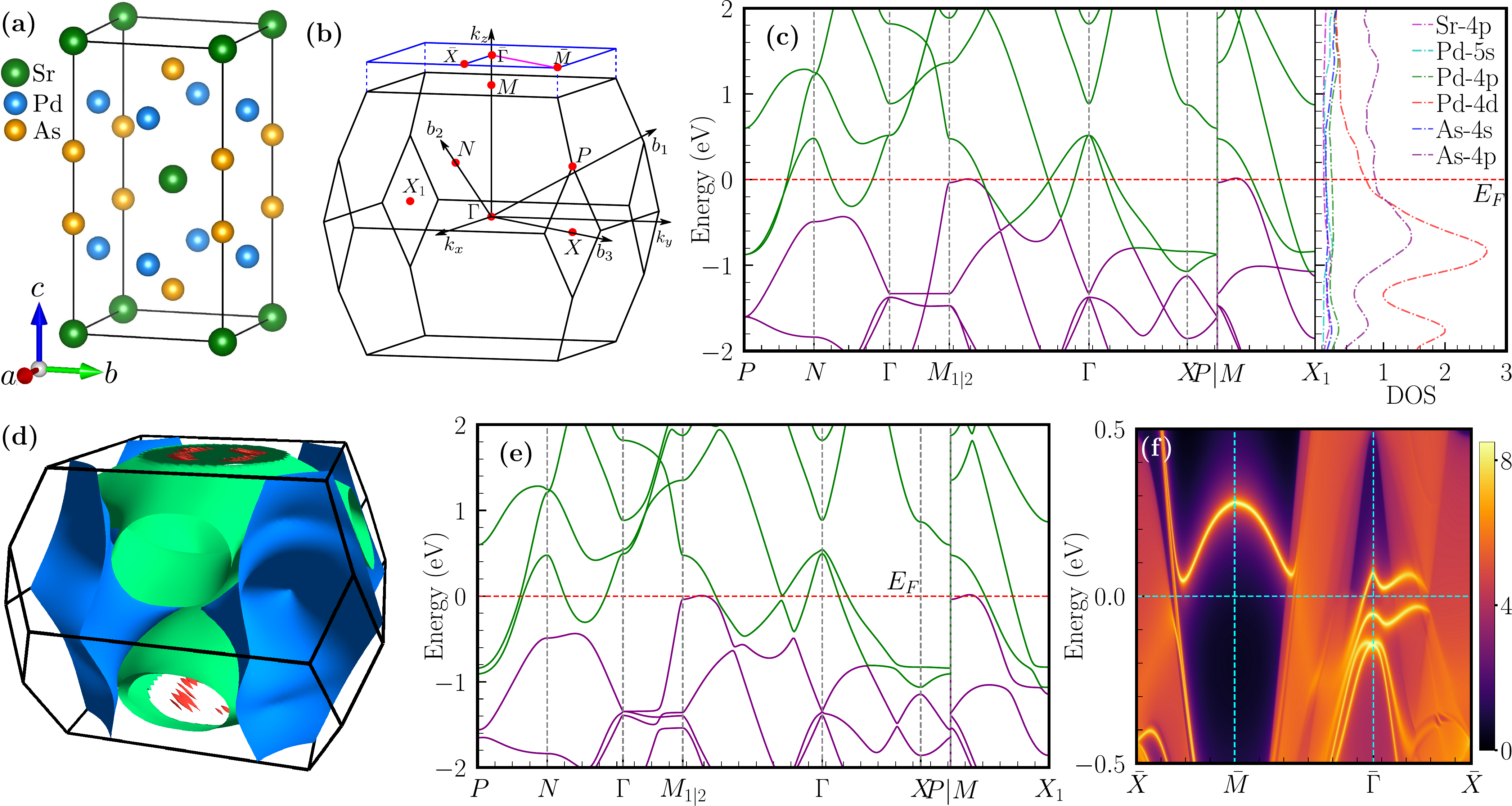}
\caption{\label{fig:band_topology} \textbf{Electronic band structure, and band topology of SrPd$_2$As$_2$:} (a) The conventional unit cell of the body-centred tetragonal crystal structure of SrPd$_2$As$_2$. (b) Bulk and $(001)$-projected surface Brillouin zones (BZs) corresponding to the body-centred tetragonal lattice. (c) Electronic band structure and orbital-resolved density of states calculated without spin-orbit coupling (SOC). (d) Combined Fermi surface sheets of SrPd$_2$As$_2$, highlighting several nearly parallel segments spanning the BZ. (e) Electronic band structure with SOC included, showing SOC-induced band splitting. (f) Surface spectral function along high-symmetry paths of the $(001)$ surface BZ, revealing multiple topological surface states.}
\end{figure*}



\section{Results}



\subsection{Electronic band structure and normal state topology}

{To obtain the electronic band structure of SrPd$_2$As$_2$, we carried out first-principles calculations using the framework of density functional theory (DFT). SrPd$_2$As$_2$ crystallizes in ThCr$_2$Si$_2$-type centrosymmetric body-centered tetragonal structure (space group $ I4/mmm$) as schematically shown in  Fig.~\ref{fig:band_topology}(a). Previous studies have established that strong hybridization between the As $4p$ and Pd $4d$ orbitals leads to the formation of robust As-Pd covalent bonds within the Pd$_2$As$_2$ layers. This pronounced orbital mixing underpins the collapsed tetragonal structure of SrPd$_2$As$_2$, as evidenced by its reduced $c/a$ ratio and shortened interlayer As-As distance \cite{anand2013superconducting,anand2012crystal} (see Appendix for structural details). The bulk and $(001)$-surface Brillouin zones (BZs) of body-centered tetragonal SrPd$_2$As$_2$ are depicted in Fig.~\ref{fig:band_topology}(b). The electronic dispersion obtained in the absence of spin-orbit coupling (SOC), presented in Fig.~\ref{fig:band_topology}(c), exhibits three doubly degenerate, strongly dispersive bands intersecting the Fermi energy, giving rise to both electron- and hole-like pockets and underscoring the intrinsically multiband character of SrPd$_2$As$_2$. The low-energy electronic states are mainly derived from Sr-$4p$, Pd-$4d$, and Pd-$4p$ orbitals, in order of decreasing contribution, corroborated by the projected density of states presented in Fig.~\ref{fig:band_topology}(c) and in agreement with earlier electronic structure studies of SrPd$_2$As$_2$~\cite{Karaca2016,shein2014}.} Introducing SOC leads to a clear splitting of the bands and opens a continuous direct gap between the valence (purple) and conduction (green) sectors in the vicinity of $E_F$, as shown in Fig.~\ref{fig:band_topology}(e). Despite this, SrPd$_2$As$_2$ remains metallic because the strongly dispersive low-energy bands continue to generate sizable electron and hole pockets. Furthermore, the centrosymmetric nature of the crystal ensures, through Kramers' theorem, that all bands remain at least two-fold degenerate across the entire Brillouin zone and SOC only induces minor deformations in the Fermi surfaces. The resulting Fermi surfaces, shown in Fig.~\ref{fig:band_topology}(d) without SOC, consists of multiple sheets. Note that all the three Fermi surface sheets are "open" or have necks at the two "poles". These specific Fermi surface geometries are critical in governing the material's low temperature thermodynamic properties.

\begin{figure*}[!htb]
\includegraphics[width=\textwidth]{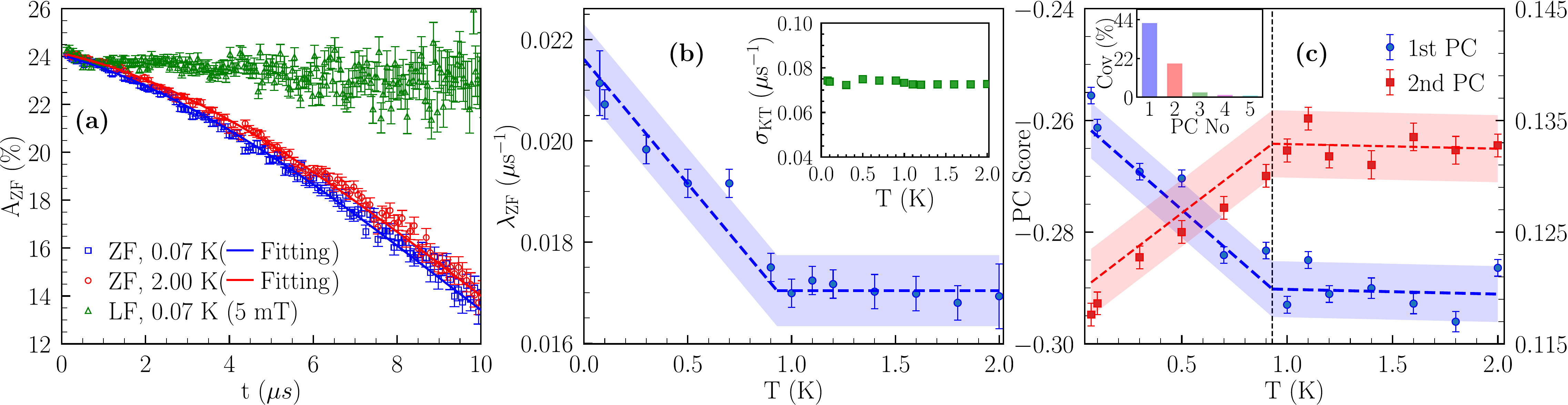} 
\caption{\textbf{Time-reversal symmetry breaking superconductivity in SrPd$_{2}$As$_{2}$:} (a) The $\mu$SR asymmetry spectra $ A_{\rm ZF}(t)$ of SrPd$_{2}$As$_{2}$ collected in zero field in superconducting state (0.07~K) and normal state (2~K) along with the fits by Eq.~(\ref{eq:MuSR_ZF}). The longitudinal field $\mu$SR spectra collected in $H = 5$~mT at temperature $T= 0.07$~K is also shown. (b) The muon spin relaxation rate $\lambda_{\rm ZF}(T)$. The dashed line is a guide to eye. Inset: Gaussian Kubo-Toyabe relaxation rate $\sigma_{\rm KT} (T)$. (c) Temperature evolution of the first and second principal component scores obtained from principal component analysis of the $\mu$SR spectra. The inset shows the percentage of covariance captured by the different principle components. Shaded regions and dashed lines are guides to the eye.}         
\label{fig:ZF}
\end{figure*}

The presence of a continuous direct gap between the conduction and valence bands throughout the BZ in SrPd$_2$As$_2$ enables a topological classification analogous to that of a fully gapped insulator~\cite{Schoop2015,Nayak2017,Ortiz2020}. {To elucidate its topological character, we determine the $\mathbb{Z}_2$ invariant by examining the evolution of hybrid Wannier charge centers (WCCs) throughout the BZ~\cite{Yu2011,Soluyanov2011}. The computed WCC evolution yields a $\mathbb{Z}_2$ index of $(1;111)$, thereby identifying SrPd$_2$As$_2$ as a $\mathbb{Z}_2$ topological metal.}

{To gain further insight into the topological character of SrPd$_2$As$_2$, we compute the surface spectral function by projecting the bulk electronic structure onto the (001) surface. The calculated spectrum exhibits multiple topologically nontrivial surface states crossing the Fermi level, including a prominent surface Dirac cone at $\bar{\Gamma} (\sim -0.14~\text{eV})$, consistent with the nontrivial $\mathbb{Z}_2$ invariant, as shown in Fig.~\ref{fig:band_topology}(f). These robust surface states provide a promising platform for realizing topological superconductivity through proximity coupling to the bulk superconducting condensate.}

\subsection{Time-reversal symmetry breaking from zero-field $\mu$SR}

The time-dependent muon asymmetry spectra of SrPd$_{2}$As$_{2}$ obtained from the zero-field (ZF) $\mu$SR measurements  are shown in Fig.~\ref{fig:ZF}(a) for representative temperatures $T=0.07$~K ($T< T_c$, superconducting state) and 2~K ($T> T_c$, normal state). The $\mu$SR spectra collected in superconducting and normal states show different muon spin relaxation rates which is more clear from the $\lambda_{\rm ZF}(T)$ plot in Fig.~\ref{fig:ZF}(b).

We fitted the ZF-$\mu$SR asymmetry spectra $ A_{\rm ZF}(t)$ of SrPd$_{2}As_{2}$ using the following damped Gaussian Kubo-Toyabe (GKT) function  \cite{hillier2022muon,bhattacharyya2018brief,mckenzie2013positive}
\begin{equation}
\label{eq:MuSR_ZF}
 A_{\rm ZF}(t)=A_0\, G_{\rm KT}(t) \,{\rm e}^{-\lambda_{\rm ZF} t} + A_{\rm BG},
\end{equation}
where, $A_{0}$ is the muon asymmetry at $t=0$ and $A_{\rm BG}$ is the non-relaxing sample holder contribution to asymmetry. $G_{\rm KT}(t)$ represents the Gaussian Kubo-Toyabe term, which is expressed as~\cite{Hayano1979,bhattacharyya2020quantum}
\begin{equation}
\label{eq:KT}
 G_{\rm KT}(t)=\left[\frac{1}{3}+\frac{2}{3}\left(1-\sigma_{\rm KT}^2 t^2 \right){\rm e}^{ -\sigma_{\rm KT}^2 t^2/2}\right]
\end{equation}
where $\sigma_{\rm KT}$ is related to the Gaussian distribution of static field $H_{\mu}$ arising from nuclear dipolar moments, given by $\sigma_{\rm KT} = \gamma_{\mu} H_{\mu}$ with $\gamma_{\mu} /2\pi$ = 135.53 MHz/T.

Our fits of the $ A_{\rm ZF}(t)$ by damped GKT according to Eq.~(\ref{eq:MuSR_ZF}) are shown by solid curves in Fig.~\ref{fig:ZF}(a). The $\lambda_{\rm ZF}(T)$ and $\sigma_{\rm KT}(T)$ for 0.07~K $\le T \le$ 2~K obtained from fitting of ZF-$\mu$SR spectra are presented in Fig.~\ref{fig:ZF}(b). While $\sigma_{\rm KT}$ remains almost constant [inset in Fig.~\ref{fig:ZF}(b)], $\lambda_{\rm ZF}$ shows a $T$-dependent behavior below $T_{c}$. A clear increase in the $\lambda_{\rm ZF}$, that sets in precisely at $T_{c}$, is evident from Fig.~\ref{fig:ZF}(b) .  The increasing $\lambda_{\rm ZF}$ attests the appearance of spontaneous internal magnetic field below $T_{c}$, and hence a TRS broken superconducting state in SrPd$_{2}$As$_{2}$.

In order to verify that the increase in $\lambda_{\rm ZF}$ is intrinsic and not caused by extrinsic effects (such as dilute magnetic impurities, trapped flux, or secondary phases) we also performed $\mu$SR measurements in longitudinal field. We note that a very small longitudinal field of 5~mT fully decouples the muons spins from the relaxation channel associated with the spontaneous internal field [see Fig.~\ref{fig:ZF}(a)] which negates the extrinsic effects to be the origin of the increase in $\lambda_{\rm ZF}$ below $T_{c}$, and confirms the appearance of static or quasistatic spontaneous magnetic field at $T < T_{c}$ that leads to the TRS breaking in the SC state of SrPd$_{2}$As$_{2}$. 

From the $\lambda_{\rm ZF}(T)$ in Fig.~\ref{fig:ZF}(b) the relaxation rate is found to change by $\Delta \lambda_{\rm ZF} \approx 0.004~\mu{\rm s}^{-1}$ which is equivalent to a spontaneous magnetic field $H_{\rm TRSB}$ of strength 0.005~mT. This value of $H_{\rm TRSB}$ for SrPd$_{2}$As$_{2}$ is much lower than the corresponding values of $H_{\rm TRSB} =0.055(5)$~mT in CaPd$_{2}$As$_{2}$ \cite{aarti2024time}, and 0.034(2)~mT in CaPd$_{2}$Ge$_{2}$ \cite{anand2023time} but is comparable to the values of 0.006~mT for ${\rm La_7Pd_3}$ \cite{mayoh2021evidence}, 0.008~mT for ${\rm La_7Rh_3}$ \cite{singh2020time} and 0.009~mT for Zr$_{3}$Ir \cite{shang2020time}. In the case of SrPtAs the $H_{\rm TRSB}$ is found to be $\approx 0.005$~mT (from $\mu$SR data measured at ISIS Facility) and  $\approx 0.007$~mT (from $\mu$SR data measured at PSI) \cite{SrPtAs}.

To sensitively track subtle temperature-dependent changes in the $\mu$SR spectra, we employ principal component analysis (PCA), an unsupervised machine-learning technique for identifying dominant trends in high-dimensional data [Fig.~\ref{fig:ZF}(c)]. The measured asymmetry spectra are decomposed into orthogonal principal components (PCs) capturing the leading variance, with the corresponding PC scores revealing systematic temperature evolution of the $\mu$SR signal. PCA has proven effective in detecting TRS breaking superconducting transitions in systems such as LaNiGa$_2$, and LaNi$_{1-x}$Cu$_x$C$_2$, as well as magnetic ordering in BaFe$_2$Se$_2$O~\cite{tula2021machine}. While highly sensitive to subtle spectral changes, PCA does not by itself identify the underlying relaxation channels and therefore complements conventional $\mu$SR analysis.

We next perform a joint principal component analysis of SrPd$_2$As$_2$ together with a set of known TRS breaking superconductors~\cite{tula2021machine}. The scree plot shown in the inset of Fig.~\ref{fig:ZF}(c) demonstrates that the first two PCs capture the dominant fraction of the total covariance in the dataset. Uncertainties in the corresponding PC scores were estimated by propagating the experimental errors of the measured asymmetry functions~\cite{tula2021machine}. As shown in Fig.~\ref{fig:ZF}(c), the temperature dependence of the PC scores exhibits a pronounced change in the first PC for $T \lesssim T_c$, signaling the onset of spontaneous time-reversal-symmetry breaking in SrPd$_2$As$_2$. The second principal component also exhibits a discernible, though weaker, variation across the superconducting transition.

\begin{figure*}[!htb]
\includegraphics[width=\textwidth]{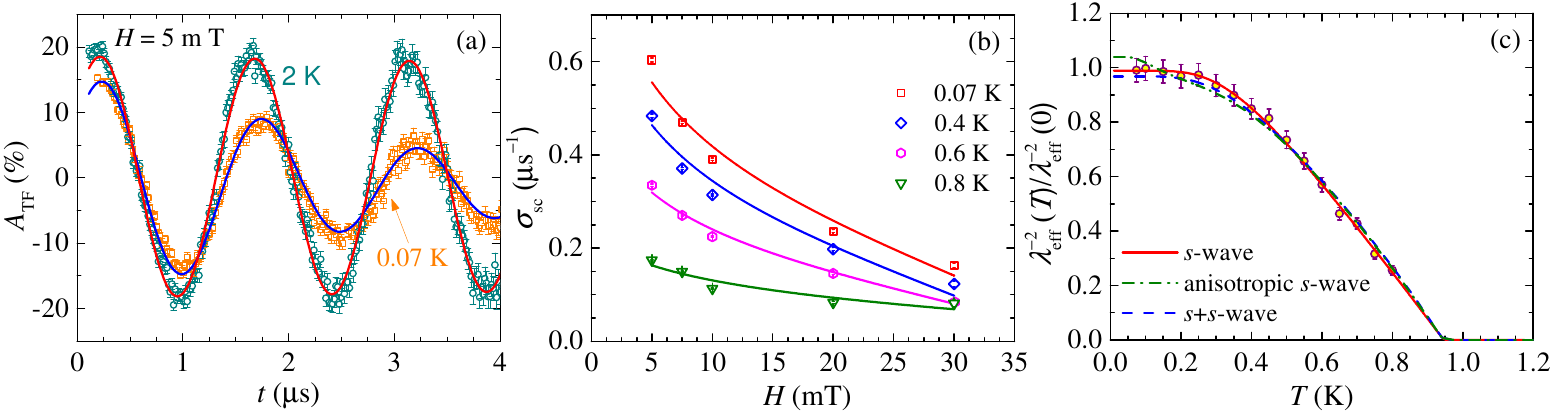} 
\caption{\textbf{Type-II bulk superconductivity with a single isotropic energy gap in SrPd$_{2}$As$_{2}$:} (a) The $\mu$SR asymmetry spectra $A_{\rm TF}(t)$ of SrPd$_{2}$As$_{2}$ collected in transverse magnetic field $H = 5$~mT (field-cooled) at 0.07~K and 2~K along with the fits by Eq.~(\ref{eq:MuSR_TF}). (b) $H$ dependence of $\sigma_{\rm sc}$ obtained from the fits of the TF-$\mu$SR spectra collected in $H =5$--30~mT along with the fits (solid lines) according to Brandt relation, Eq.~(\ref{eq:sigma_sc}). (c) Normalized effective penetration depth $\lambda_{\rm eff}$ obtained using  Eq.~(\ref{eq:sigma_sc}) plotted as $\lambda_{\rm eff}^{-2}(T) / \lambda_{\rm eff}^{-2}(0)$ along with the fits by Eq.~(\ref{eq:lambda_eff}) for the isotropic single-gap $s$-wave model, anisotropic $s$-wave model and two-gap ($s+s$)-wave models.}         
\label{fig:TF}
\end{figure*}

\subsection{Superconducting gap structure from transverse-field $\mu$SR}

The superconducting order parameter/gap structure of SrPd$_2$As$_2$ has been probed by transverse-field $\mu$SR measurements, which also allowed us to estimate the penetration depth. The TF $\mu$SR asymmetry spectra $A_{\rm TF}(t)$ of SrPd$_{2}As_{2}$ collected at representative temperatures of 0.07~K (superconducting state) and 2~K (normal state) are shown in Figs.~\ref{fig:TF}(a) measured in field-cooled condition in a transverse field of 5~mT. The Fourier transform of the $A_{\rm TF}(t)$ spectra [Appendix Fig.~\ref{fig:TF1}] clearly show 
the presence of inhomogeneous field distribution and vortex state in the superconducting state (at 0.07~K), that classifies SrPd$_{2}As_{2}$ as a type-II superconductor as also inferred from value of Ginzburg-Landau parameter $k_{\rm GL} = 7.8$ in the previous study \cite{anand2013superconducting}.

The TF-$\mu$SR asymmetry spectra of SrPd$_{2}$As$_{2}$ are well described by a Gaussian oscillatory function with an oscillatory background~\cite{panda2024probing, adroja2021pairing,adroja2021evidence}
\begin{equation}
\label{eq:MuSR_TF}
\begin{split}
 A_{\rm TF} (t) &= A'_0 \cos\left(\omega t + \phi \right) {\rm e}^{ -\sigma_{\rm TF}^2 t^2/2} \\
  & \hspace{1cm} + A_{\rm BG} \cos\left(\omega_{\rm BG} t + \phi \right).
\end{split}
\end{equation}
where $\sigma_{\rm TF}$ is the Gaussian relaxation rate,  and $A'_0$ and A$_{\rm BG}$ are the initial asymmetries (at $t=0$) of sample and silver sample-holder, respectively. The muon precession frequency $\omega$ = $\gamma_{\mu}H_{\rm int}$  provides the estimate of the internal magnetic field $H_{\rm int}$ sensed by muons, and $\phi$ is the initial phase of the muon spin precession. Similarly, $\omega_{\rm BG}$ = $\gamma_{\mu}H_{\rm int,BG}$, with $H_{\rm int,BG}$ denoting the internal field associated with the background signal (i.e. muons stopping on the Ag-holder), which also estimates the value of the applied TF.

The fits of $A_{\rm TF} (t)$ by the function in Eq.~(\ref{eq:MuSR_TF}) are shown in Fig.~\ref{fig:TF}(a). The  Gaussian relaxation rate $\sigma_{\rm TF}$ consists of two contributions, $\sigma_{\rm TF}^2$ = $\sigma_{\rm sc}^2$ + $\sigma_{\rm nm}^2$, where $\sigma_{\rm sc}$ represents the superconducting contribution on account of the inhomogeneous field distribution in the vortex state, and $\sigma_{\rm nm}$ represents the nuclear dipolar moment contribution. While the normal state $\sigma_{\rm TF}(T)$ (at $T>T_c$) contains only nuclear contribution (that provides us the estimate of $\sigma_{\rm nm}$), the SC state $\sigma_{\rm TF}(T)$ (at T $\le$ $T_{c}$) contains contributions from both $\sigma_{\rm nm}$ and $\sigma_{\rm sc}$. The value of  $\sigma_{\rm sc}$ is extracted by subtracting off the value of  $\sigma_{\rm nm}$ from  $\sigma_{\rm TF}$ according to $\sigma_{\rm TF}^2$ = $\sigma_{\rm sc}^2$ + $\sigma_{\rm nm}^2$. The $T$ dependence of $\sigma_{\rm TF}$ obtained from the fits of TF-$\mu$SR asymmetry spectra at different $T$ by Eq.~(\ref{eq:MuSR_TF}) and the extracted $\sigma_{\rm sc}$ for various applied fields are shown in Appendix Fig.~\ref{fig:TF-sigma}. The field dependence of $\sigma_{\rm sc}$ is shown in Fig.~\ref{fig:TF}(b) for few representative temperatures.

The $T$ and $H$ dependent $\sigma_\mathrm{sc}$ of SrPd$_{2}$As$_{2}$ was further analyzed according to Brandt's model \cite{brandt1988magnetic,brandt2003properties,panda2019probing}.
\begin{equation}
\begin{split}
\sigma_\mathrm{sc} & = \frac{4.83 \times 10^{4}}{\lambda_{\rm eff}^{2}} (1-H_\mathrm{ext}/H_\mathrm{c2})  \\ 
& \hspace{1cm}\times [1+1.21\left(1-\sqrt{{(H_\mathrm{ext}/H_\mathrm{c2}})} \right)^{3}]
\end{split}
\label{eq:sigma_sc}
\end{equation}
which allowed us to obtain the effective magnetic penetration depth $\lambda_{\text{eff}}$, subject to the condition that $H_{\text{ext}}/H_{c2} \leq 0.25$ and  Ginzburg-Landau parameter $k_{\rm GL} \ge 5$, where $\sigma_\mathrm{sc}$ is expressed in the unit of $\mu s^{-1}$ and $\lambda_{\text{eff}}$ in nm.  In Eq.~(\ref{eq:sigma_sc}),  $H_{c2}$ is the upper critical field and $H_\mathrm{ext}$ is the applied transverse field.

For SrPd$_2$As$_2$, the upper critical field is $H_{c2} = 70$~mT~\cite{anand2013superconducting}, which is significantly higher than the $H_{\rm ext}$ applied in the TF $\mu$SR measurements. Together with a large Ginzburg--Landau parameter $\kappa_{\rm GL} = 7.8$, this justifies the use of the Brandt equation to analyze the $H$ dependence of $\sigma_{\rm sc}$. The temperature-dependent effective penetration depth $\lambda_{\rm eff}(T)$ (which provides the estimate of superfluid density) was obtained by fitting the $\sigma_{\rm sc}(H)$ data shown in Fig.~\ref{fig:TF}(b), with the corresponding fits indicated by solid curves. The resulting normalized penetration depth (superfluid density), $\lambda_{\rm eff}^{-2}(T)/\lambda_{\rm eff}^{-2}(0)$, is presented in Fig.~\ref{fig:TF}(c).

The $T$ dependence of the normalized superfluid density was analyzed using the standard expression~\cite{prozorov2006magnetic,bhattacharyya2021electron,bhattacharyya2024exploring}
\begin{equation}
\frac{\lambda_{\rm eff}^{-2}(T,\Delta)}{\lambda_{\rm eff}^{-2}(0)}
= 1 + \frac{1}{\pi} \int_{0}^{2\pi} \int_{\Delta(T,\varphi)}^{\infty}
\frac{\partial f}{\partial E}
\frac{E\,{\rm d}E\,{\rm d}\varphi}{\sqrt{E^{2}-\Delta^{2}(T,\varphi)}} ,
\label{eq:lambda_eff}
\end{equation}
which provides direct information about the $T$- and angle $\varphi$-dependent superconducting gap structure. Here, $f = [1+\exp(E/k_{\rm B}T)]^{-1}$ is the Fermi--Dirac distribution function and $E$ denotes the quasiparticle energy.

The superconducting energy gap $\Delta(T,\varphi)$ is assumed to depend on both temperature and momentum according to $\Delta(T,\varphi) = \Delta(0)\,\delta(T/T_c)\,g(\varphi)$~\cite{annett1990symmetry,pang2015evidence}. Here, $\Delta(0)$ is the zero-temperature gap magnitude, $\delta(T/T_c)$ describes its temperature evolution, and $g(\varphi)$ encodes the angular dependence of the gap. The temperature-dependent factor is approximated as $\delta(T/T_c) \approx \tanh\!\left[1.82\{1.018(T_c/T - 1)\}^{0.51}\right]$~\cite{carrington2003magnetic}. For an isotropic $s$-wave gap, $g(\varphi)=1$, while for an anisotropic $s$-wave gap it takes the form $g(\varphi)=|1+\cos(4\varphi)|/2$~\cite{annett1990symmetry,pang2015evidence,bhattacharyya2019investigation}.

The experimental $\lambda_{\rm eff}^{-2}(T)/\lambda_{\rm eff}^{-2}(0)$ data were analyzed using three models: (i) a single isotropic $s$-wave gap, (ii) a single anisotropic $s$-wave gap, and (iii) a two-gap isotropic $(s+s)$-wave model. The corresponding fits using Eq.~(\ref{eq:lambda_eff}) are shown in Fig.~\ref{fig:TF}(c) and the fit parameters are listed in Appendix Table~I. Among these, the single isotropic $s$-wave model provides the best description of the data (see SM for details) implying a thermodynamically fully gapped superconducting state, i.e., low-energy quasiparticle excitations are absent on the Fermi surfaces of SrPd$_2$As$_2$. From this fit, we obtain a superconducting gap value $\Delta(0) = 0.127(2)$~meV, corresponding to a gap ratio $2\Delta(0)/k_{\rm B}T_c = 3.13(5)$ with $T_c = 0.94$~K. This value is smaller than the weak-coupling BCS limit of 3.53, but slightly larger than the value of 3.0 (corresponding to $\alpha = 1.5$) inferred from heat-capacity measurements~\cite{anand2013superconducting}. The same analysis yields an effective penetration depth $\lambda_{\rm eff}(0) = 321(17)$~nm, which is larger than the values $\lambda_{\rm eff}(0) = 130$~nm estimated from $H_{c2}$ and $\lambda_{\rm eff}(0) = 170(70)$~nm obtained from tunnel diode resonator (TDR) measurements~\cite{anand2013superconducting}. A similar enhancement of $\lambda_{\rm eff}(0)$ extracted from $\mu$SR relative to TDR has also been observed in CaPd$_2$As$_2$~\cite{aarti2024time}, and is likely attributable to the use of polycrystalline samples in $\mu$SR measurements, in contrast to single-crystal samples used for TDR studies.

The zero-temperature penetration depth further allows an estimate of the superconducting carrier density $n_s \approx m^{*}c^{2}/(4\pi e^{2}\lambda_{\rm eff}^{2}(0))$ following the approach in Refs.~\cite{hillier1997classification, anand2011specific,adroja2021pairing}. For SrPd$_2$As$_2$, the effective mass $m^{*} = 1.44\,m_e$, where $m_e$ is the free-electron mass ~\cite{anand2013superconducting}. Using $\lambda_{\rm eff}(0) = 321$~nm, we estimate $n_s = 3.94 \times 10^{26}$~m$^{-3}$. From this carrier density, the Fermi temperature is obtained via $k_{\rm B}T_{\rm F} = (\hslash^{2}/2m^{*})(3\pi^{2}n_s)^{2/3}$, yielding $T_{\rm F} = 1580$~K.

The ratio $T_c/T_{\rm F}$ provides a useful metric for classifying superconductors within the Uemura scheme~\cite{uemura1989universal,uemura1991classifying}. Conventional superconductors typically satisfy $T_c/T_{\rm F} \leq 0.001$, whereas for unconventional superconductors $0.01 \leq T_c/T_{\rm F} \leq 0.1$. For SrPd$_2$As$_2$, the values $T_c = 0.94$~K and $T_{\rm F} = 1580$~K yield $T_c/T_{\rm F} = 5.95 \times 10^{-4}$, placing it within the conventional BCS regime according to this classification. Notably, the TRS-breaking superconductors CaPd$_2$As$_2$ and CaPd$_2$Ge$_2$ also fall within the same conventional region of the Uemura plot~\cite{aarti2024time,anand2023time} (see Appendix Fig.~\ref{fig:Uemura}).

\section{Discussions}
The estimated superconducting-state parameters and the Uemura classification of SrPd$_2$As$_2$ initially suggest a phonon-mediated, single-band, weak-coupling BCS type superconductivity. However, these indicators are accompanied by a notable anomaly: a reduced specific-heat jump at $T_c$, with $\Delta C_{e}/\gamma_{n}T_{c} \approx 1.03$~\cite{anand2013superconducting}. This corresponds to a coupling parameter $\alpha = \Delta(0)/k_{\rm B}T_{c} = 1.50$, which is significantly lower than the standard BCS value $\alpha_{\rm BCS} = 1.764$~\cite{anand2013superconducting}. Within the framework of the $\alpha$-model, such a reduction is commonly interpreted as evidence for an anisotropic superconducting gap or multigap superconductivity~\cite{anand2013superconducting,padamsee1973quasiparticle,johnston2013elaboration}. Similar reduced $\alpha$ values have also been reported for the isostructural compounds CaPd$_2$Ge$_2$ ($\alpha = 1.62$) and CaPd$_2$As$_2$ ($\alpha = 1.58$)~\cite{anand2013superconducting,anand2014superconductivity}. Moreover, this thermodynamic indication of gap anisotropy is at odds with our $\mu$SR investigations, which reveal clear signatures of broken time-reversal symmetry (TRS) in SrPd$_2$As$_2$, similar to its Ca-based counterparts~\cite{anand2023time,aarti2024time}. This unusual combination of a reduced specific-heat jump, fully gapped behavior as inferred from TF-$\mu$SR, and spontaneous TRS breaking points toward an unconventional superconducting ground state that defies a simple classification.

\begin{figure}[!t]
\includegraphics[width=0.5\textwidth,origin=b]{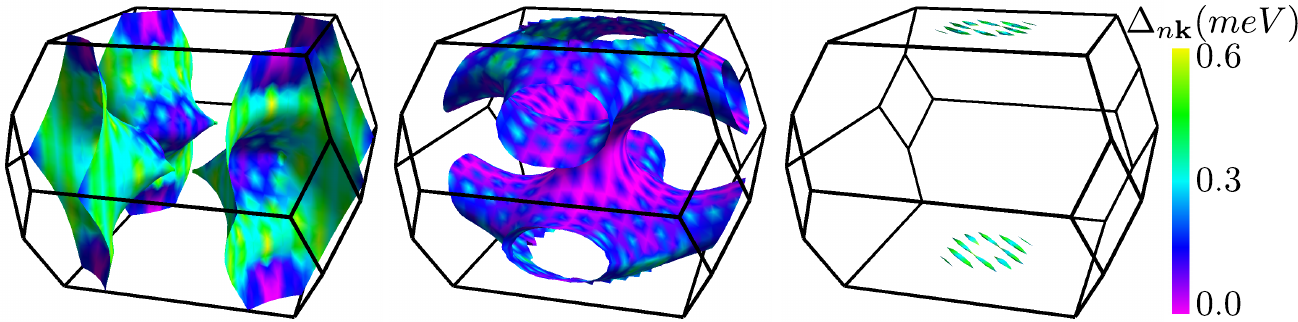}
\caption{\textbf{Momentum-dependent superconducting gap with complex nodal structure from anisotropic Migdal-Eliashberg theory:} Superconducting gap distribution on the three Fermi surface sheets of SrPd$_2$As$_2$, evaluated at T = $0.5$ K. The gap shows strong anisotropy across the different sheets and momentum directions, with multiple nodes characteristic of nodal superconductivity. Although Migdal-Eliashberg theory predicts a complex nodal structure, transverse-field $\mu$SR measurements reveal a fully gapped state. This contradiction effectively rules out a conventional phonon-mediated mechanism in SrPd$_2$As$_2$.}
\label{fig:aniso_gap} 
\end{figure}

To critically assess the expectations from a conventional electron--phonon-mediated pairing scenario, we theoretically investigated the possible phonon-mediated superconducting ground state of SrPd$_2$As$_2$. After establishing dynamical stability of SrPd$_2$As$_2$ (Appendix Fig.~\ref{fig:phonon}), we evaluated the electron--phonon coupling strength and estimated the superconducting transition temperature using the Allen--Dynes-modified McMillan formula~\cite{Allen_1975}, obtaining $T_c = 1.27$~K, which significantly overestimates the experimental value of $0.94$~K (see Appendix for details). Furthermore, solving the fully anisotropic Migdal-Eliashberg equations~\cite{Margine_2013} yields a strongly momentum-dependent superconducting gap $\Delta(\mathbf{k})$, varying from $0.0$ to $0.6$~meV, with multiple nodes distributed across the Fermi surface sheets, as shown in Fig.~\ref{fig:aniso_gap}. This theoretical prediction of phonon-mediated superconducting ground state with complex nodal structure is in direct contradiction with transverse-field $\mu$SR measurements, which establish a thermodynamically fully gapped superconducting state having no low-energy quasiparticle excitations on the Fermi surfaces of SrPd$_2$As$_2$.

In addition, possible TRS breaking superconducting states in SrPd$_2$As$_2$ are severely constrained by symmetry considerations: as a centrosymmetric material, SrPd$_2$As$_2$ allows only pure singlet or pure triplet pairing states, thereby excluding mixed-parity superconductivity. Moreover, the recently proposed possible loop-supercurrent mechanism for TRS breaking in fully gapped superconductors is also forbidden in SrPd$_2$As$_2$, since its primitive unit cell contains only one formula unit with a single symmetrically distinct site for each atomic species~\cite{ghosh2021time}, whereas this mechanism requires multiple symmetry inequivalent sites. Taken together, the coexistence of a fully gapped superconducting state and spontaneous TRS breaking in SrPd$_2$As$_2$ provides compelling evidence for an unconventional superconducting ground state.

 To reconcile the apparent contradiction of isotropic single gap superconductivity with the observation of TRS-breaking in SrPd$_2$As$_2$, we employ a symmetry-based analysis within the Ginzburg-Landau (GL) framework~\cite{ghosh2021Recent,Sigrist1991,annett1990symmetry}. The crystal structure of SrPd$_2$As$_2$ has symmorphic space group symmetry $I4/mmm$ (No. $139$), with the corresponding point group $D_{4h}$. It is paramagnetic in the normal state. Hence, the normal state symmetry group $\mathcal{G} = G \otimes U(1) \otimes \mathcal{T}$, where $G$ includes the crystal point-group operations together with spin rotations, $U(1)$ represents global gauge symmetry, and $\mathcal{T}$ denotes TRS, thus fully determine the possible symmetry allowed uniform superconducting instabilities. Being centrosymmetric SrPd$_2$As$_2$ can have either purely singlet or triplet superconducting state even though the effect of SOC is significant. The $D_{4h}$ point group contains 8 one-dimensional (1D) irreducible representations (irreps) -- four of even parity and four of odd parity -- as well as 2 two-dimensional (2D) irreps, $E_g$ (even parity) and $E_u$ (odd parity). A TRS-breaking superconducting state can arise only from these two 2D irreps and such a superconducting state will necessarily be unconventional since it requires breaking of additional crystalline symmetries.

\begin{figure}[!t]
\includegraphics[width=0.95\columnwidth,origin=b]{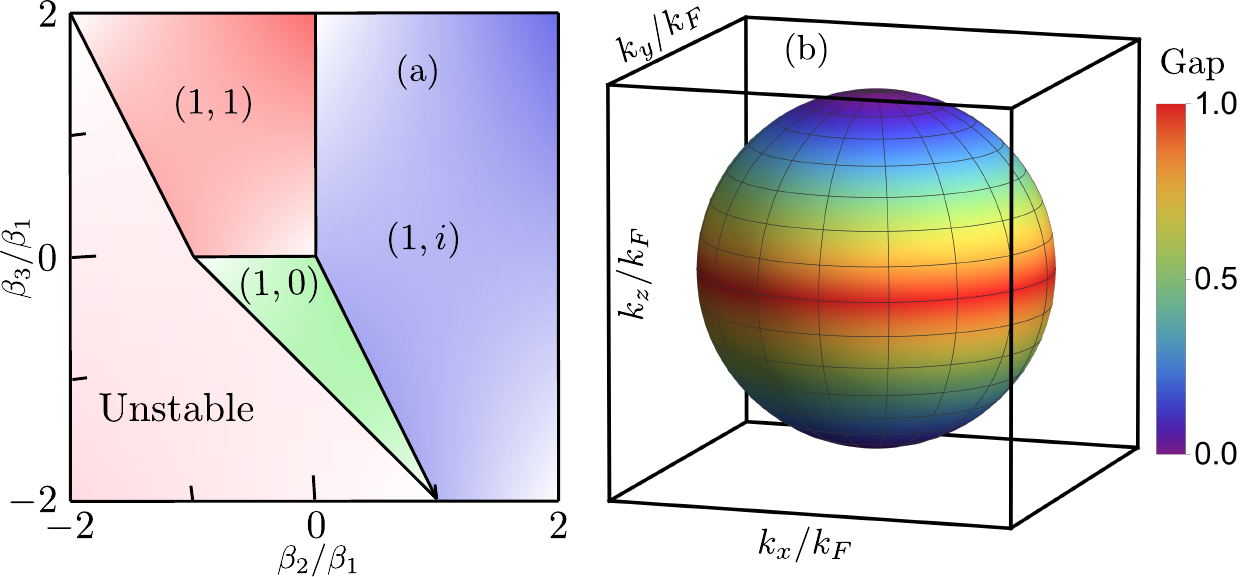}
\caption{\textbf{Nonunitary triplet superconducting ground state in SrPd$_2$As$_2$:} (a) Ginzburg-Landau phase diagram showing the symmetry-allowed two-component order parameters $(\eta_1,\eta_2)$ corresponding to the two-dimensional irreducible representations of the point group $D_{4h}$. A time-reversal-symmetry-breaking superconducting state is stabilized for $(\eta_1,\eta_2)=\tfrac{1}{\sqrt{2}}(1,i)$ over an extended region of the parameter space. (b) Polar plot of the minimum quasiparticle excitation gap on a spherical Fermi surface for the nonunitary triplet superconducting state which spontaneously breaks TRS and have 2 point nodes located at the north and south poles.}
\label{fig:GL} 
\end{figure}

We therefore focus exclusively on the $E_g$ and $E_u$ irreps and construct the possible TRS-breaking superconducting order parameters compatible with the symmetry of the system and construct the fourth-order GL invariants associated with these 2D irreps of $D_{4h}$. For a two-component order parameter $\pmb{\eta}=(\eta_1,\eta_2)$, the symmetry-allowed quartic terms of the GL free energy take the form
\begin{equation}
    f_4 = \beta_1 (|\eta_1|^2+|\eta_2|^2)^2 + \beta_2 |\eta_{1}^2 + \eta_{2}^2|^2 + \beta_3 (|\eta_1|^4+|\eta_2|^4)\,. \label{eqn:GL_fenergy}
\end{equation}
\noindent Here, $\beta_1$, $\beta_2$, and $\beta_3$ are material-dependent GL coefficients. Minimizing the free energy yields three symmetry-allowed superconducting states: $(\eta_1, \eta_2) = (1,0)$, $\frac{1}{\sqrt{2}}(1,1)$, and $\frac{1}{\sqrt{2}}(1,i)$. Notably, there exists a substantial region of parameter space in which the chiral state $\frac{1}{\sqrt{2}}(1,i)$ is energetically favored, as shown in Fig.~\ref{fig:GL}(a). This state spontaneously breaks TRS at $T_c$ due to the nontrivial phase difference between the two components of the order parameter. Assuming strong SOC, the symmetry-allowed even-parity TRS-breaking order parameter in the pseudospin singlet channel for the $E_g$ irreducible representation is therefore given by $\Delta(\mathbf{k})=\Delta_0 k_z (k_x + ik_y)$ with $\Delta_0$ as a momentum-independent amplitude. This corresponds to a chiral $d$-wave pseudospin singlet state. The odd-parity pseudospin triplet superconducting state belonging to the $E_u$ irrep is characterized by a $\mathbf{d}$-vector of the form $\mathbf{d}(\mathbf{k})=[Ak_z,iAk_z,B(k_x+ik_y)]$. Here, $A$ and $B$ are real, material-dependent constants that are independent of $\mathbf{k}$. We also note that $\mathbf{d}(\mathbf{k}) \times \mathbf{d}^*(\mathbf{k})=2iAk_z(Bk_x \hat{x}-Bk_y\hat{y}-Ak_z\hat{z})$, which is generically nonzero, demonstrating that this superconducting state is a nonunitary chiral $p$-wave pseudospin triplet state. 

To determine the nature of the superconducting gap for the two candidate TRS-breaking chiral states, we begin by computing the Bogoliubov quasiparticle excitation spectrum on a generic single-band spherical Fermi surface using Bogoliubov-de Gennes mean-field theory~\cite{Sigrist1991,Ghosh2020}. The chiral $d$-wave pseudospin singlet order parameter has the energy gap $|\Delta_0 ||k_z| (k^2_x + k^2_y)^{1/2}$, which features two point nodes at the ``north and south'' poles, as well as an additional equatorial line node at $k_z=0$. Thermodynamically, the low-temperature behavior of this state would be dominated by the line node, as it contributes a significantly larger low-energy density of states than the point nodes. In contrast, the triplet state exhibits an energy gap of $[B^2 (k^2_x + k^2_y) + 2 A^2 k^2_z - 2 |A||k_z| \{A^2 k^2_z + B^2 (k^2_x + k^2_y)\}^{1/2}]^{1/2}$ which is shown in Fig.~\ref{fig:GL}(b) choosing $A$ and $B$ to be unity for example. This state thus hosts only two point nodes located at the poles of the spherical Fermi surface, lacking the equatorial line node present in the singlet case.

The preceding analysis, based on a generic Fermi surface, can be directly adapted to the multi-band system SrPd$_2$As$_2$ by incorporating the specific momentum dependence of the gap on each Fermi surface sheet while neglecting interband pairing. As shown in Fig.~\ref{fig:band_topology}(d), all the three Fermi surface sheets in SrPd$_2$As$_2$ are geometrically ``open'', featuring necks at the two poles. Consequently, the odd-parity nonunitary triplet state, which possesses point nodes at the poles only, would effectively avoid these nodes in this specific geometry since the polar point nodes do not intersect the Fermi surface sheets. Thus, the odd-parity chiral nonunitary triplet state would behave effectively as a fully gapped superconductor, with low-temperature Fermi-surface-averaged thermodynamic properties (such as the superfluid density measured by TF-$\mu$SR) indistinguishable from those of a conventional isotropic gap system and will be stabilized in SrPd$_2$As$_2$. However, unlike a conventional superconductor, this state naturally accounts for the spontaneous time-reversal symmetry breaking observed in SrPd$_2$As$_2$. In contrast, the even-parity chiral singlet state will still have the equatorial line node intersecting two of the Fermi surface sheets of SrPd$_2$As$_2$. The low-temperature thermodynamic properties for this state will therefore show a power-law behavior characteristic of a superconducting gap with line nodes and thus this state is incompatible with the phenomenology of SrPd$_2$As$_2$.

The proposed bulk nonunitary triplet superconducting state in SrPd$_2$As$_2$ can constitute a possible intrinsic realization of topological superconductivity since SrPd$_2$As$_2$ is a $\mathbb{Z}_2$ topological metal, which is equivalent to a doped strong topological insulator~\cite{Hao_2011,Hsieh_2012}. Moreover, its topological surface states that are well-separated from the bulk bands and disperse across the Fermi level can acquire a unique surface superconducting gap via the proximity effect from the bulk condensate. This establishes SrPd$_2$As$_2$ as a promising platform for realizing topological superconductivity. The resulting coexistence of distinct bulk and surface superconducting gaps can, for example, be experimentally resolved using techniques such as point-contact Andreev reflection spectroscopy~\cite{mehta2024topological} or high-resolution angle-resolved photoemission spectroscopy (ARPES)~\cite{Yang2023}. The Fermi surface topologies of SrPd$_2$As$_2$ predicted from DFT calculations can be verified using de Haas-van Alphen measurements on single crystals for example.

\section{Conclusions}
By integrating microscopic probes with first-principles calculations and symmetry analysis, we establish that the superconducting state of the $\mathbb{Z}_2$ topological metal SrPd$_2$As$_2$ is consistent with nonunitary triplet pairing. Zero-field $\mu$SR provides clear evidence for spontaneous time-reversal symmetry (TRS) breaking, while the transverse-field response indicates a fully gapped superconducting state -- a combination not captured within standard phonon-mediated pairing as shown by our anisotropic Migdal-Eliashberg calculations. We resolve this apparent contradiction by demonstrating that the material’s symmorphic $D_{4h}$ symmetry and open Fermi surface topology stabilize an odd-parity triplet ground state whose symmetry-imposed nodes are located in momentum-space regions devoid of electronic states. This allows the system to exhibit an effectively isotropic thermodynamic response while naturally breaking TRS.

These results establish SrPd$_2$As$_2$ as a clean platform for studying nonunitary triplet pairing in topological metals, where topology and crystalline symmetry strongly constrain the superconducting order parameter. More broadly, our work provides a concrete mechanism by which TRS breaking can arise in systems that appear fully gapped in bulk probes, suggesting a possible intrinsic route toward topological superconductivity. At the same time, the emergence of such a state in a system with relatively weak electronic correlations and weak electron–phonon coupling highlights an open question regarding the microscopic origin of the pairing interaction, motivating further theoretical and experimental investigation.

\section*{Acknowledgments}
DTA thanks the Royal Society of London for the Newton Advanced fellowship funding between UK and China, and the International Exchange funding between UK and Japan. DTA also thanks EPSRC UK (Grant number EP/W00562X/1) for funding and the CAS for PIFI Fellowship. AB expresses gratitude to the Science and Engineering Research Board for the CRG Research Grants (CRG/2020/000698 \& CRG/2022/008528) and CRS Project Proposal at UGC-DAE CSR (CRS/2021-22/03/549). We thank ISIS Facility for providing beam time  RB2410594 and RB2510560~\cite{RB24, RB25}. SKG acknowledges financial support from Anusandhan National Research Foundation (ANRF), erstwhile Science and Engineering Research Board (SERB), Government of India via the Startup Research Grant: SRG/2023/000934. DS and SKG utilized the \textit{Andromeda} server at IIT Kanpur for numerical calculations.   

\begin{figure}
\includegraphics[width=0.9\columnwidth]{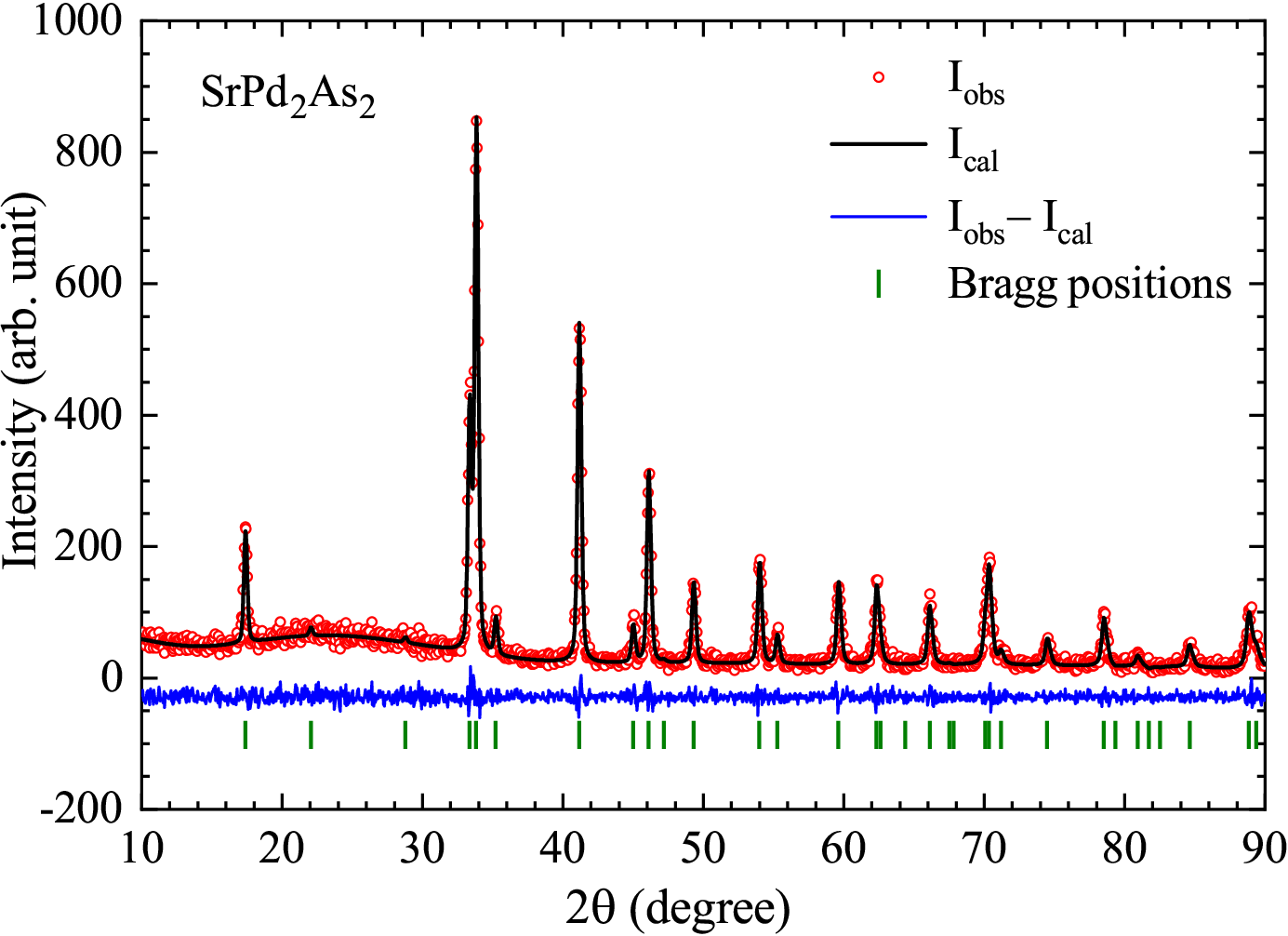}.
 \caption{The room-temperature powder x-ray diffraction pattern of SrPd$_{2}$As$_{2}$. The black lines represent the Rietveld refinement profile for ThCr$_{2}$Si$_{2}$-type body-centered tetragonal (space group $I4/mmm$) structure. The vertical bars indicate the positions of the Bragg reflections, while the blue curve at the bottom displays the difference between the observed and calculated intensities}.       
\label{fig:XRD-Fig}
\end{figure}

\section*{Appendix}

\subsection{X-ray Diffraction}

The powder x-ray diffraction data of  SrPd$_{2}$As$_{2}$ collected at room temperature is shown in  Fig.~\ref{fig:XRD-Fig} along with the Rietveld refinement profile.  The refinement of XRD pattern confirmed the ThCr$_2$Si$_2$-type body-centered tetragonal structure (space group $ I4/mmm$) of SrPd$_{2}$As$_{2}$ and the single phase nature of the sample. The refined lattice parameters were found to be $a = b = 4.3817(2)$ \AA\ and $c =10.1803(6)$ \AA, and the $z$ coordinate of As atom along the $c$-axis was found to be $z_{As} =0.3647(4)$. These values of refined lattice parameters and atomic coordinate agree well with the values obtained for the SrPd$_2$As$_2$ single crystals \cite{anand2013superconducting}.

\begin{figure*}[!htb]
\includegraphics[width=1.0\textwidth,origin=b]{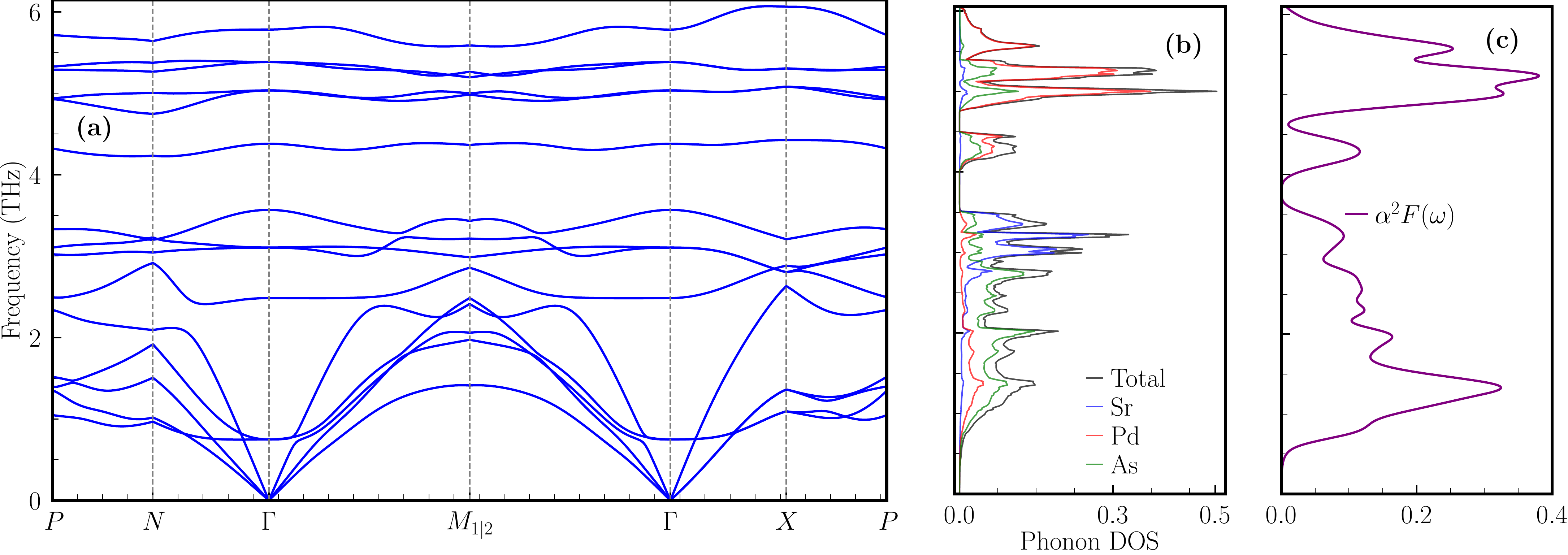}
\caption{{Phonon dispersion, density of states, and Eliashberg spectral function of SrPd$_{2}$As$_{2}$:} (a) Calculated phonon dispersion curves, showing the absence of any unstable (imaginary) modes. (b) Phonon density of states. (c) Isotropic Eliashberg spectral function $\alpha^2 F(\omega)$.}
\label{fig:phonon} 
\end{figure*}

\subsection{Phonon-Mediated superconductivity}

To examine the lattice dynamics and quantify the electron-phonon interactions in SrPd$_{2}$As$_{2}$, we carried out phonon and electron-phonon coupling calculations using density-functional perturbation theory together with the EPW~\cite{Ponce_2016,Margine_2013,Giustino_2007} framework implemented in the QUANTUM ESPRESSO package. The unit cell of SrPd$_{2}$As$_{2}$ contains five atoms, giving rise to fifteen phonon branches-three acoustic and twelve optical-as shown in \fig{fig:phonon}(a). The phonon dispersion exhibits no unstable modes, as evidenced by the absence of imaginary frequencies, confirming the dynamical stability of the crystal structure. As seen in \fig{fig:phonon}(b), the phonon modes below $2.7$ THz originate primarily from vibrations of As atoms, those in the $2.7-3.5$ THz range are mainly associated with Sr vibrations, and vibrations of Pd atoms dominate the high-frequency modes above $4$ THz. The Eliashberg spectral function $\alpha^2 F(\omega)$, shown in \fig{fig:phonon}(c), exhibits prominent peaks in both the low- and high-frequency regions with comparable magnitudes, indicating that electron-phonon coupling is particularly strong in these frequency ranges relative to the intermediate vibrational modes.

The integrated electron-phonon coupling constant $\lambda = 2 \int (\alpha^2 F(\omega)/\omega)\,d\omega$ is $\lambda = 0.65$, placing SrPd$_{2}$As$_{2}$ in the weak-coupling superconducting regime. The logarithmically averaged phonon frequency $\omega_{log}=exp[(2/\lambda) \int (\alpha^2 F(\omega)/\omega) \ln \omega\,d\omega]$ is $\omega_{log}=1.79$ THz for SrPd$_{2}$As$_{2}$. The McMillan formula, modified by Allen and Dynes, can be used to estimate the superconducting critical temperature based on these values~\cite{Allen_1975}:

\begin{equation}
    T_c = \frac{\omega_{log}}{1.2} exp \left[ \frac{-1.04(1+\lambda)}{\lambda(1-0.62 \mu^*)-\mu^*} \right]
\end{equation}

\noindent Here, $\mu^*$ is the Anderson-Morel Coulomb pseudopotential~\cite{Morel_1962}, which is difficult to calculate reliably from first principles. Instead, it is often inferred empirically from tunneling experiments, where typical values fall in the range $\mu^*=0.10-0.20$~\cite{McMillan_1968,Dynes_1972,Allen_1975}. Using $\mu^*=0.16$, we estimate a superconducting critical temperature of $T_c=1.27$ K, which is slightly higher than the experimentally observed value of $0.92(5)$ K.


\begin{figure}
\includegraphics[width=\columnwidth]{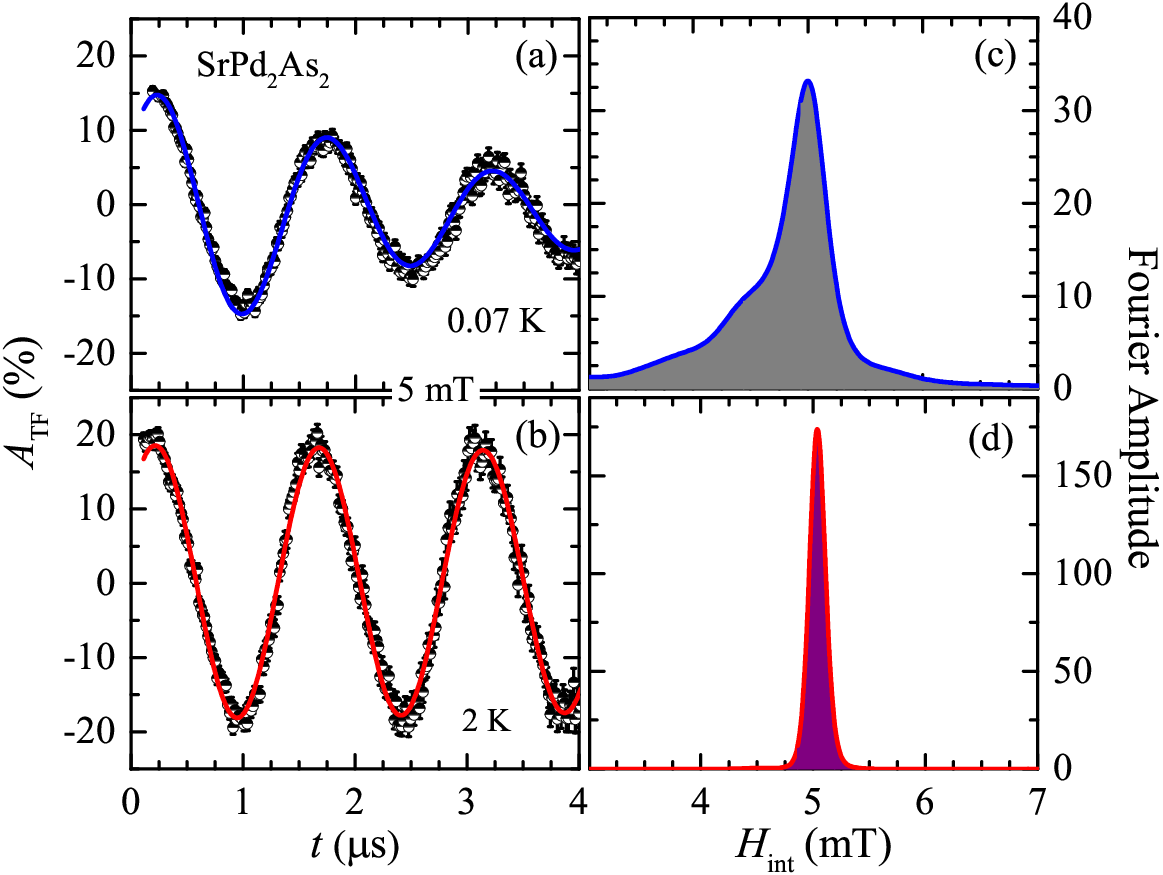}
\caption{ Time $t$ evolution of muon asymmetry $A_{\rm TF}$ spectra of SrPd$_{2}$As$_{2}$ measured in field-cooled condition in an applied transverse magnetic field $H = 5$~mT at (a) 0.07~K and (b) 2~K with their corresponding Fourier transformed maximum entropy spectra in (c) and (d), respectively.}
\label{fig:TF1}
\end{figure}

\begin{table*}[!htb]
\caption{Parameters obtained from fits to $\lambda_{\rm eff}^{-2}(T)/\lambda_{\rm eff}^{-2}(0)$ for SrPd$_2$As$_2$ using a single isotropic $s$-wave gap, a single anisotropic $s$-wave gap, and a two-gap ($s+s$)-wave model.}
\label{Table}
\centering
\setlength{\tabcolsep}{4pt}
\footnotesize
\begin{ruledtabular}
\begin{tabular}{@{} l c c c c @{}}
Model & weight & $\Delta_i(0)$ (meV) & $2\Delta/k_{\rm B}T_c$ & $\chi^2$ \\
\hline
Isotropic $s$-wave      & 1          & 0.127(2)              & 3.13(5)              & 1.30 \\
Anisotropic $s$-wave    & 1          & 0.165(1)              & 4.07(3)              & 3.95 \\
Two-gap ($s+s$)-wave    & 0.86, 0.14 & 0.140(3), 0.032(1)    & 3.453(8), 0.789(4)   & 4.06 \\
\end{tabular}
\end{ruledtabular}
\end{table*}

\subsection{Transverse field $\mu$SR}

The Fourier transform of the transverse field $\mu$SR asymmetry $A_{\rm TF}(t)$ spectra at 0.07~K and 2~K [in Figs.~\ref{fig:TF1}(a) and \ref{fig:TF1}(b)] are shown in Figs.~\ref{fig:TF1}(c) and \ref{fig:TF1}(d), respectively. A glance at Figs.~\ref{fig:TF1}(c) and \ref{fig:TF1}(d) clearly show marked differences between the Fourier transforms in superconducting state and normal state. The differences can be attributed to the inhomogeneous field distribution caused by the formation of a vortex state in the SC phase of SrPd$_{2}$As$_{2}$, as a result the muons  depolarize with different rates in SC and normal states.

\begin{figure} [t]
\includegraphics[width=0.9\columnwidth]{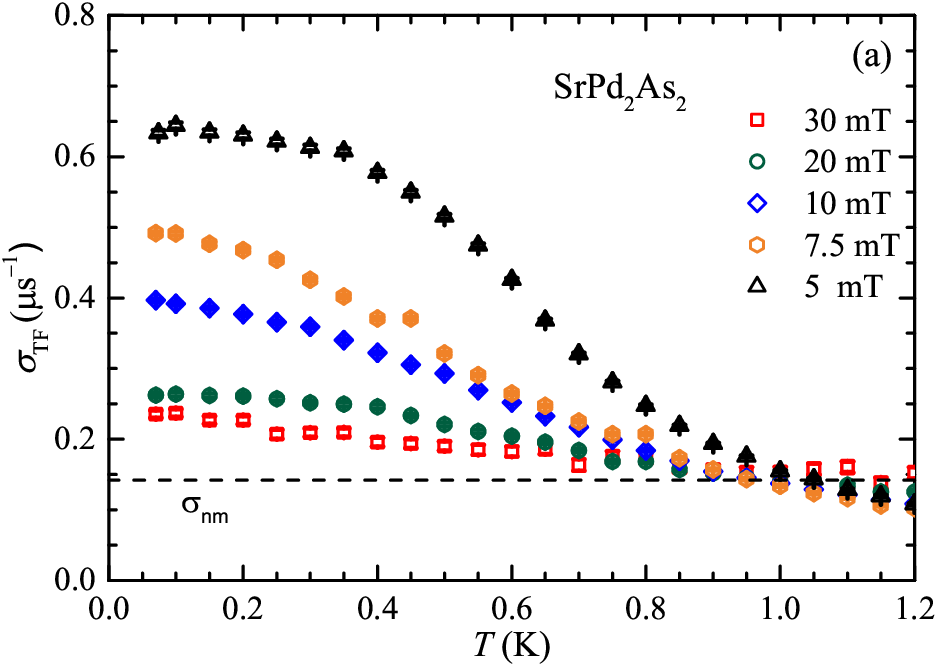}
\includegraphics[width=0.9\columnwidth]{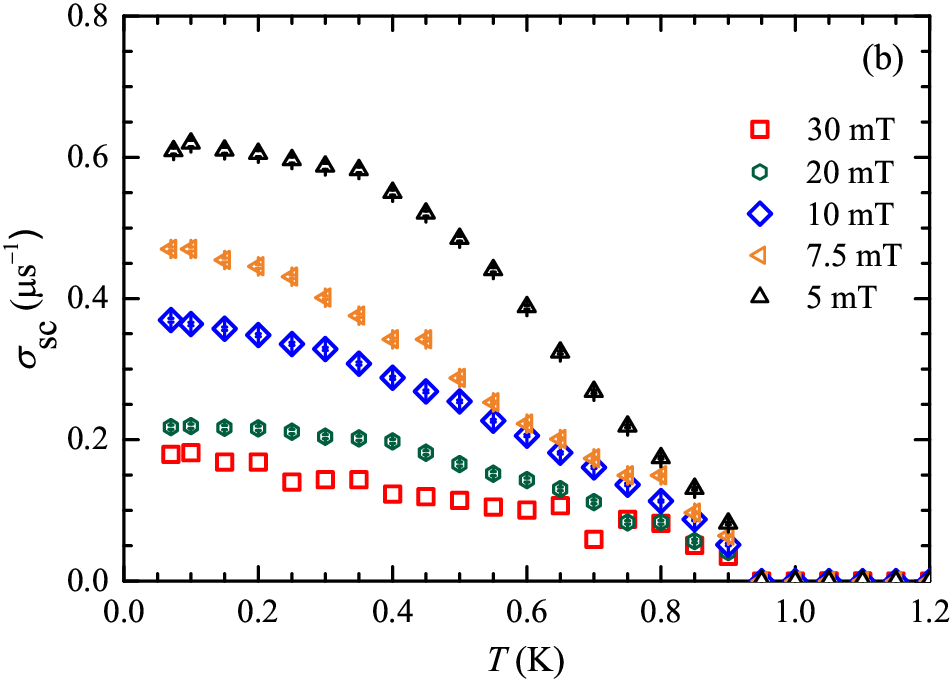}
\caption{(a) Temperature $T$ dependence of muon spin relaxation rate $\sigma_{\rm TF}$  obtained from the fits of the TF-$\mu$SR spectra of SrPd$_{2}$As$_{2}$ collected in field-cooled state with the indicated applied transverse magnetic fields $H$. (b) The superconducting contribution to muon spin relaxation rate $\sigma_{\rm sc}$ versus $T$.}         
\label{fig:TF-sigma}
\end{figure}

Figure~\ref{fig:TF-sigma}(a) shows the temperature $T$ dependence of the Gaussian relaxation rate $\sigma_{\rm TF}$ obtained from the fits of the TF-$\mu$SR spectra at various temperatures. The superconducting contribution $\sigma_{\rm sc}$ to $\sigma_{\rm TF}$ is shown in Fig.~\ref{fig:TF-sigma}(b).

We analyzed the TF-$\mu$SR data for  $\lambda^{-2}_{\text{eff}}(T)/\lambda^{-2}_{\text{eff}}(0)$ using three different models: i) single isotropic $s$-wave gap model, ii) single anisotropic $s$-wave gap model, and iii) two isotropic ($s+s$)-wave gap model. The values of the goodness of the fit parameters ($\chi^2$) and other parameters are listed in Table.~\ref{Table}. For the isotropic $s$-wave gap model we obtained superconducting gap $\Delta(0)  = 0.127(2)$~meV, corresponding to a gap ratio $2\Delta(0)/k_{\rm B} T_c = 3.13(5)$ with $T_c = 0.94$~K, which is smaller than the BCS weak-coupling limit 3.53 but a little larger than the value of 3.0 (for $\alpha= 1.5$) obtained from the analysis of the heat capacity data \cite{anand2013superconducting}. From the isotropic $s$-wave fit model we also estimate a value of $\lambda_{\rm eff}(0) = 321(17)$~nm which is higher than the value $\lambda_\text{eff}(0) = 130$~nm obtained from the value of $H_{c2}$ and $\lambda_\text{eff}(0) = 170(70)$~nm obtained from the penetration depth measurement using a tunnel diode resonator (TDR).

\begin{figure}[!t]
\includegraphics[width=\columnwidth]{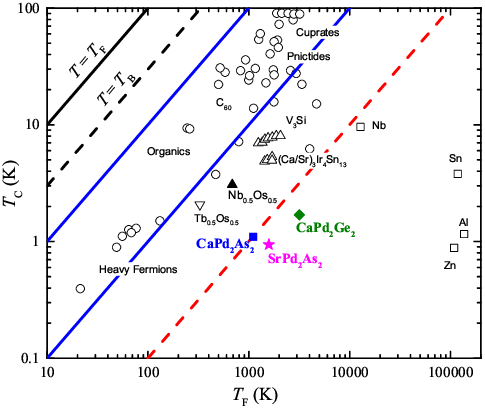}
\caption {The Uemura plot, superconducting $T_{c}$ versus Fermi temperature $T_{\rm F}$ for various types of superconductors. The positions of SrPd$_{2}$As$_{2}$, CaPd$_{2}$As$_{2}$ and CaPd$_{2}$Ge$_{2}$ on this plot classify them as the conventional superconductors.} 
\label{fig:Uemura}
\end{figure}
Figure~\ref{fig:Uemura} shows the Uemura plot, for various superconductors. According to the Uemura classification based on the values of the ratio $T_{c}$/$T_{\rm F}$, SrPd$_{2}$As$_{2}$ falls in the category of conventional superconductors as shown in Figure~\ref{fig:Uemura}. The TRS broken superconductors CaPd$_{2}As_{2}$ and CaPd$_{2}$Ge$_{2}$ also fall \cite{aarti2024time,anand2023time} in this category of conventional superconductors.

\bibliography{ref}

\end{document}